\documentclass[preprint,12pt]{elsarticle}
\usepackage{amssymb}
\usepackage[]{epsfig}
\usepackage{amsmath}
\usepackage{pstricks}
\journal{Physics Letters B}
\begin{document}
\begin{frontmatter}

\title{Double copy of spontaneously broken Abelian gauge theory}
\author{Euro Spallucci\footnote{e-mail:Euro.Spallucci@ts.infn.it} $\,$\footnote{Senior Associate}}
\address{INFN,\\
Sezione di Trieste, Trieste, Italy}
\author{Anais Smailagic\footnote{e-mail:Anais.Smailagic@ts.infn.it} $\,$\footnote{Senior Associate}}
\address{INFN, Sezione di Trieste, Italy}

\begin{abstract}

Similarity in the structure of scattering amplitudes in Yang-Mills theories and
 General Relativity led to the idea that graviton could be described as the \emph{ double copy} of a vector gauge field. In this letter we discuss a  realization
of this idea emerging directly from solutions of equations of General Relativity. A general form of the energy momentum tensor for the electric field  is derived that leads to the metric tensor in terms of the double copy of a corresponding gauge potential. 
We than use this general property to find the double copy of  spontaneously broken
scalar electrodynamics. The result is a \emph{screened} Reissner-Nordström-like  metric. 
When the horizon radius and the Compton wavelength of the massive photon become comparable
the black hole becomes a quantum object. An exact solution of the horizon
wave equation is found and the corresponding energy spectrum is described.
It turns out that highly excited states show a characteristic string-like  behavior.
\end{abstract}
\end{frontmatter}

\section{Introduction}

Expressing gravity in the language analogous to that of gauge theories
is still an uncompleted task. On this path, there are at least three main
obstacles to be surmounted. The most evident one derives from the fact that
gauge field is a one-index Lorentz vector while graviton is a two-index symmetric tensor. The first difference is the respective geometric meanings of these fields. Gauge fields are connections appearing inside 
covariant derivatives of Yang Mills theory while it is Christoffel symbol which is the connection in the
gravitational covariant derivatives. The gravitational  connection is a composite function of  metric and its derivatives derivatives and not a fundamental field. On the other hand, graviton field describes deviation from flatness in the \emph{metric tensor}. 

The second difference is that Lagrangians for gauge theories are always \emph{quadratic}
in the field strength while the Hilbert-Palatini action in General Relativity is
linear in the curvature of the metric. Attempts to gauge Lorentz, Poincar\`e or
other groups do not lead straightforwardly to General Relativity but to more general theories like Riemann-Cartan geometry.\\
Finally, the gauge coupling constant is dimensionless ( in natural units $c =
1 , \hbar = 1$) while the Newton constant has dimension of length squared  $G_N= L^2_{Pl}/2$.\\ In spite of many attempts, starting with the seminal papers by Utiyama \cite{Utiyama} and Kibble  \cite{Kibble:1961ba}  up to the conjectured duality
between strings in  $AdS_5$ and boundary SUSY $SU\left(\,N\,\right)$  gauge theories \cite{Maldacena:1997re}, the
 satisfactory unified description of gauge theories and gravity has not been reached.\\
A  new approach to the  connection between gravity and gauge theories was recently proposed based on
 the structural similarity among perturbative scattering amplitudes
in Yang-Mills (YM) theory and  General Relativity. We postpone
technical detail of this approach to the next section and only mention here
the final conclusions which can be summarized in the statement  ”~gravity is the square
of YM theory~” \cite{Bern:1999vk,Bern:2010yg,Bern:2010ue,Bern:2014zja,Bern:2014dlu,Bern:2015gsa,Bern:2017yxu}  or, equivalently, that graviton is the double copy of a  gauge field. Extending this conjecture beyond the scattering amplitude level leads
to the picture in which the neutral Schwarzschild field can be described as the double
copy (DC) of the Coulomb potential \cite{Monteiro:2014cda,Monteiro:2015bna,White:2017mwc,Bahjat-Abbas:2017htu,Bahjat-Abbas:2020cyb,Alkac:2021bav,Gonzo:2021drq}. In other words, a neutral black hole solution has been designed as the DC of a point-like Coulomb electric field once  correspondence  between  respective coupling constants is assumed.
However, this type of  correspondence is not  compatible with other solutions of General Relativity equations. It is 
known  that the Einstein-Maxwell equations with a massive charged particle as the source admit the
Reissner-Nordström metric as an exact solution. In this solution, however, mass and charge are \emph{distinct, unrelated parameters}.\\
In this letter we would like to go beyond DC scheme and describe  Reissner-Nordström graviton  as a "~new~" double copy (NDC) of  the Coulomb  potential.
The paper is organized as follows: in Section(2) we review the distinctive properties of the Kerr-Schild gauge and briefly explain how it helps to describe the graviton as the DC of a gauge field. To pave the way towards
a NDC relation between the gauge potential and the graviton, we first describe a derivation of the Reissner-Nordström solution in this framework.


 In Section(3) we concentrate on the core part of the paper and describe the gravitational field as double copy of the Abelian-Higgs model.
 The resulting metric has a Reissner-Nordström-like form 
except that  the charge term has a dumping Yukawa-like factor. Below the screening length induced by the photon mass, quantum effects become dominant inducing horizon uncertainty fluctuations.
We solve the horizon wave equation and determine the energy spectrum. Highly excited 
states reveal an interesting relationship with quantum strings.

\section{Graviton as the double copy of the photon}
\label{dc}
Asymptotically flat static/stationary  metrics can be cast in  a  simple form in the so called Kerr-Schild gauge. In this gauge, the tensor structure
is determined by a null vector~\footnote{$l_\mu$ is null both respect the Minkowski background metric and the full line element  $\eta_{\mu\nu}l^\mu l^\nu=g_{\mu\nu}l^\mu l^\nu=0$  } $l_\mu$ and a single scalar \emph{potential} $\phi$. The resulting line element reads    

\begin{equation}
ds^2 =ds^2_{\small{Mink} }- 2\phi\, l_\mu\, l_\nu \,dx^\mu\, dx^\nu\ ,\quad 
l_\mu\equiv \left(\, 1\ , \frac{\vec{x}}{\vert\vec{x}\vert} \,\right)\ . 
\label{uno}
\end{equation}

In the \emph{spherical gauge} the line element (\ref{uno}) takes the usual form

\begin{equation}
ds^2 =-\left(\, 1 + 2\phi\,\right) dt^2 +\left(\, 1 + 2\phi\,\right)^{-1} dr^2 +r^2\left(\, d\theta^2 + \sin^2 \theta\, d\phi^2 \,\right)
\ .
\end{equation}
 The reason why we referred to $\phi$ as a scalar "~potential~" becomes evident 
 in the simplest case of a point-like mass $m$ where $\phi = -mG_N/r$ is nothing else but the Newtonian potential.\\
The metric tensor in (\ref{uno})  defines  the graviton field $h_{\mu\nu}$
as

\begin{equation}
\kappa \, h_{\mu\nu}\equiv -2 \phi \, l_\mu\, l_\nu \ ,\qquad \kappa \equiv \sqrt{8\pi G_N}\ . \label{due}
\end{equation}
The constant $\kappa$ is introduced in order to provide canonical dimension to the bosonic field $h_{\mu\nu}$.
From equation (\ref{due}) we see that the tensor structure of the graviton is given by the product of two copies of
the null vector $l_\mu$. In this description the graviton is the\emph { double copy} of  $l_\mu$.
\\
The "miracle" of the Kerr-Schild gauge is that the Einstein field equations reduce to the single
Poisson equation  for the unknown field $\phi$ in flat space:

\begin{equation}
R^\mu{}_{\nu}-\frac{1}{2}\delta^\mu {}_\nu R =8\pi G_N \, T^\mu{}_\nu \longrightarrow \nabla^2\phi 
= -8\pi G_N \left(\, T^0{}_0-\frac{1}{2} T^\mu{}_\mu\,\right)\ . \label{tre}
\end{equation}
We stress that this is an \emph{exact} result and \emph{no weak field expansion} is involved.\\
In this letter we are interested in solving equation (\ref{tre}) with the energy momentum tensor for
an Abelian gauge theory.\\
Perturbative gluon-gluon  and graviton-graviton scattering amplitudes show  a surprising structural similarity
suggesting that the graviton is the double copy of the gauge vector field or, alternatively, the gauge field is the
single copy of the graviton

\begin{equation}
h_{\mu\nu}\sim \phi\, l_\mu\,\l_\mu   ,\qquad A_\mu \sim \phi \, l_\mu\ .
\end{equation}

Following this line of reasoning one finds a \emph{duality} DC relationship between the Newton and Coulomb potential
by matching  the respective coupling constants
\begin{eqnarray}
\phi =-\frac{G_N m}{r} && \longleftrightarrow \phi =-\frac{e}{4\pi r} \ ,\label{dcdual1}\\
G_N m && \longleftrightarrow \frac{e}{4\pi }\ .\label{dcdual2}
\end{eqnarray}
Translating the above relations in a relativistic language one concludes that the Schwarzschild metric is the 
DC of the Coulomb potential, provided the electric charge is replaced with the effective gravitational coupling as shown above.\\
 On the other hand, one also knows  that the metric produced by a static, massive point-like charge is the  Reissner-Nordstr\"om space-time. \\
In DC  approach the Reissner-Nordstrom geometry has been only reconstructed perturbatively up to $O\left(\, G_N^2\,\right)$ by introducing a uniform ball of charged dust as a source in  gravity equations \cite{Sardelis:1973em}.
However,  a smooth limit to a point source requires a redefinition of a physical mass in terms of the electric charge
and the radius of the ball.\\
In alternative to the DC approach we shall find that one can solve the
Einstein equations leading to the general result  which is quadratic in the gauge potential and linear in the Newton one. 
 In our approach, mass and charge remain distinct parameters but the graviton can still be described as
a  NDC of the photon. We shall summarize our results in the concluding Sect.(\ref{end}).
\\
The first part of our discussion is aimed to clarify the sense in which the Reissner-Nordstr\"om  geometry can be seen
as the NDC of a Coulomb field.

\subsection{NDC derivation of the Reissner-Nordstr\"om solution}

In  \cite{Easson:2021asd} and \cite{Godazgar:2021iae} the Reissner-Nordstro\"om metric was recovered 
in the framework of Weyl double copy using spinorial language.  Here we shall obtain the same metric in the framework of
General Relativity in Kerr-Schild gauge. Let us start from the Lagrangian for an Abelian gauge potential
$A_\mu$

\begin{equation}
L\left[\, A\,\right]= -\frac{1}{4}\,\partial_{[\,\mu}  \,  A_{\nu\,]}\, \partial^{[\,\mu}\, A^{\nu\,]}-q J^\mu A_\mu\ .
\label{max}
\end{equation}
The vector current $J^\mu$ is the field source and $q$ is the electric charge.\\
 We assume the source to be vanishing everywhere except in $r=0$. 
 The general form of the electromagnetic field energy momentum tensor reads \footnote{ We adopt the definition:
 \begin{equation*}
T_{\mu\nu}=-2\frac{\delta \mathcal{L}}{\delta g^{\mu\nu}} +g_{\mu\nu} \mathcal{L}\ .
\end{equation*}
}

\begin{equation}
T_{\mu\nu}=F_{\mu\lambda}\, F_\nu{}^\lambda -\frac{1}{4}\eta_{\mu\nu} F_{\alpha\beta}F^{\alpha\beta}
\end{equation}

It is immediate to see that $T^\mu{}_\mu=0$ and we need only $T^0{}_0$ in the r.h.s. of the Poisson equation.\\
The electric field generated by a \emph{static}  charge $q$ is time independent.  Thus, both  $\partial_0 A_0=0$ and 
$\partial_0 F_{0m}=0$ and no magnetic field is present $F_{mn}=0$. Furthermore, if the charge is point-like the current density reads 
 $J^\mu \equiv \delta^\mu{}_0 \, \delta\left(\, \vec{r}\,\right)$ and solving Maxwell equations one finds
the Coulomb potential $A^0=-q/4\pi r$ and the radial electric field $E^r=-q/4\pi r^2$.
For $T^0{}_0$ we find:
\begin{equation}
T^0{}_0=F^0{}_k\, F_0{}^k -\frac{1}{2}\,F_{0m}\,F^{0m}     
\end{equation}

By definition the  electric field is $E^m\equiv F^{0m}$  and one can  write $T^0{}_0$ as

\begin{equation}
T^0{}_0=-\frac{1}{2}\, \vec{E}\cdot \vec{E}=-\frac{e^2}{32\pi^2 \, r^4} \ .
\label{edensity}
\end{equation}

In General Relativity textbooks (\ref{edensity}) is inserted in the Einstein field equations leading to the Reissner-Nordstr\"om solution in spherical coordinates. \\
Here we introduce a simpler and general way for solving the problem also when the electric field is not divergence free and  the r.h.s. is of the form

\begin{equation}
\nabla^2\phi = 4\pi G_N\left(\, \vec{E}\cdot \vec{E}-A^0 \nabla\vec{E}\,\right)
\label{peq}
\end{equation}

We notice that

\begin{eqnarray}
\nabla^2 \left(\, A^0 A^0\,\right)  && = 2\left(\,\nabla A^0\cdot \nabla^0 A^0 + A^0 \nabla^2 A^0 \,\right)\ ,\nonumber\\
&&= 2\left(\,\vec{E}\cdot \vec{E} - A^0 \nabla \vec{E} \,\right)\ . \label{a2}
\end{eqnarray}

Identity (\ref{a2}) allows to write (\ref{peq})  as
\begin{equation}
\nabla^2\phi = 2\pi G_N\nabla^2 \left(\, A^0 A^0\,\right) \ .
\label{peq1}
\end{equation}
There is no need of further steps to solve (\ref{peq1}) and write the general solution as
\begin{eqnarray}
&& \phi =\phi_0 + 2\pi G_N \, A^0 A^0 \ ,\label{sol}\\ 
&& \phi_0= c_1 + \frac{c_2}{r}\ , \label{homo}
\end{eqnarray}

$\phi_0$ is the solution of the homogeneous Poisson equation $\nabla^2 \phi_0=0$. The two integration constants
$c_1$, $c_2$ are fixed by the \emph{physical} conditions defining the problem we are considering: \\
i) asymptotic flatness implies $ c_1=0$;\\
ii) the source is a static charge which must be supported by a \emph{massive} particle. The mass of the
particle generates its own gravitational field, thus $ c_2=- G_N m$.\\
The complete solution and the graviton field read

\begin{eqnarray}
&&\phi= -\frac{G_N m}{r}+ 2\pi G_N \, A^0 A^0\ ,
\label{stot}\\
&& \mathcal{A}_\mu\equiv A^0 l_\mu\ ,\quad A^0 = -\frac{e}{4\pi r}\ ,\label{nullgamma}\\
&&\kappa\, h_{\mu\nu}= \frac{2G_N m}{r} l_\mu l_\nu- 4\pi G_N \,\mathcal{A}_\mu \, \mathcal{A}_\nu\ . \label{grav}
 \end{eqnarray}

Equation (\ref{grav}) displays  the Newtonian and Coulombic components of the graviton.  It is also evident that each contribution
is again the double copy of $l_\mu$ but each term has a \emph{different potential function} in front. The gravitational
potential appears \emph{once} in the neutral part of graviton while the electric potential  enters \emph{twice} in the 
charged term.\\
The important conclusion of the first part of this paper is that \emph{every time  the energy density 
can be written in the form} (\ref{edensity}) \emph{the Einstein-Maxwell equations are solved by} (\ref{stot}).\\

\section{Higgs electrodynamics}
\label{ssb}
The Higgs mechanism is a milestone of spontaneously broken gauge theories. It is, therefore,  interesting to look for the corresponding gravitational double copy.  In  the true vacuum the expectation value of the Higgs field is non-vanishing, i.e. $v_0\neq 0$, and the   photon field acquires mass. The simplest model exhibiting Higgs mechanism is scalar QED
with a symmetry breaking potential. String theory derived effective quantum field theories involving additional dilaton and axion fields are beyond the purpose of this letter. Here we shall study the simplest case.\\
The  dynamics of the heavy photon 
is described by the effective non-local Lagrangian\footnote{It may be interesting to note that the non-local Lagrangian
(\ref{nlh})  can be derived in the path integral formalism. By expanding the original Lagrangian around the true vacuum up to quadratic terms in modulus of the complex scalar field  one obtains a Gaussian integral for the phase (i.e. the Goldstone boson) which can be integrated away. The resulting one-loop effective Lagrangian for $A_\mu$ is (\ref{nlh}).
It is worth to note that integrating away the Goldstone boson, instead of imposing the unitary gauge, preserves gauge invariance even in the presence of the photon mass. } :

\begin{equation}
L\left[\,A\,\right]= -\frac{1}{4}\,F_{\mu\nu}\left(\, 1-\frac{\mu^2}{\partial^2} \,\right)F^{\mu\nu}    - q J^\mu A_\mu\ ,\qquad
\mu^2=e^2 v_0^2\ . \label{nlh}
\end{equation}
 The current    density $J^\mu= \delta^\mu{}_0\delta\left(\, \vec{r}\,\right) $ leads to the single 
 field equation 
 \begin{equation}
\left(\, 1 -\frac{\mu^2}{\nabla^2}\,\right) \nabla^2 A^0 =  q\,\delta\left(\, \vec{r}\,\right)
\label{A} 
\end{equation}
which is solved by a Yukawa-type potential

  \begin{equation}
A^0 = -\frac{q}{4\pi r} e^{-\mu r}\ .\\
\end{equation}
The corresponding  electric field is
\begin{equation}
E^r= -\frac{q}{4\pi r^2} \left(\, 1 + \mu r\, \right) \, e^{-\mu r}\ .
\end{equation}

In the non-trivial Higgs vacuum electrostatic interaction is short-range and  the wavelength of the massive photon represents the screening length.\\
By proceeding as in the previous section we find the following energy momentum tensor
\begin{equation}
T_{\mu\nu}=F_{\mu\lambda}\left(\, 1- \frac{\mu^2}{\partial^2}\,\right) F_\nu{}^\lambda +\eta_{\mu\nu} \mathcal{L}\ .
\end{equation}

The resulting electrostatic energy density reads

\begin{equation}
T^0{}_0=-\frac{1}{2}\,\vec{E}\left(\, 1- \frac{\mu^2}{\nabla^2}\,\right)\cdot\vec{E} \ . \label{t11}
\end{equation}

Using the identity

\begin{eqnarray}
&&\vec{E}\frac{1}{\nabla^2}\vec{E}=  - A^0 A^0\ ,\\
\end{eqnarray}

we see that the non-local term in (\ref{t11}) can be written as  $\mu^2 A^0 A^0$. As $A^0$ is a Yukawa type
potential satisfying a massive Kelin-Gordon type equation we have

\begin{equation}
\mu^2 A^0 A^0 = A^0 \nabla^2 A^0= -A^0 \nabla \vec{E}
\end{equation}

The general discussion of the previous section allows us to write the gravitational potential as
\begin{equation}
\phi = -\frac{mG_N}{r} + \frac{q^2G_N}{8\pi^2} \frac{e^{-2\mu r}}{r^2}\ ,
\end{equation}
where $m$ is the mass of the test charge $q$ not to be confused with the photon mass $\mu$.\\
Thus, the broken Abelian gauge theory has its gravitational dual copy in the form  of a \emph{screened} Reissner-Nordstr\"om metric

\begin{eqnarray}
ds^2=&& ds^2_{\small{Mink} }-2\left[\, -\frac{mG_N}{r} + \frac{q^2G_N}{8\pi^2}\frac{ e^{-2\mu r}}{r^2} \,\right]   \, l_\mu\, l_\nu \,dx^\mu\, dx^\nu\ ,\nonumber\\
=&&-\left[\, 1 -\frac{2mG_N}{r} + \frac{q^2G_N}{4\pi^2} \frac{e^{-2\mu r} }{r^2}  \,\right] dt^2 \nonumber\\
&&+\left[\, 1 -\frac{2mG_N}{r} + \frac{q^2G_N}{4\pi^2} \frac{e^{-2\mu r}}{r^2}\,\right]^{-1} dr^2
+r^2 d\Omega^2\ . \label{yuk}
\end{eqnarray}

One observes that, in agreement with the short range nature of the electrostatic interaction in the Higgs vacuum, the charge contribution
 exponentially vanishes at a distance greater than the characteristic screening length ($r> 1/2\mu$) (\ref{yuk}) and the metric looks like the  Schwarzschild one.\\
 The horizon equation for this geometry is non-linear and cannot be solved in a closed form

\begin{equation}
r_h^2 -2mG_N r_h +\alpha_q G_N e^{-2\mu r_h}=0\ ,\qquad \alpha_q \equiv \frac{q^2}{4\pi}\ .
\end{equation}
However, we can profit from the presence of two different length scales to get approximate solutions. In fact, we have:
\begin{enumerate}
	\item the usual Schwarzschild radius $r_h=2mG_N$ \\
	\item the Compton wavelength $\lambda_A=1/2\mu $ of the (double copy) massive photon contribution to the graviton.\\
\end{enumerate}
When $r_h > 1/2\mu$, or $m >1/4G_N \mu$,  (\ref{yuk}) describes a classical  black hole of radius and temperature
given by
\begin{eqnarray} 
&& r_+ \simeq  2m G_N \left(\, 1 -\frac{\alpha_q}{4m^2 G_N} e^{-4\mu m G_N}\,\right)\ ,\\
&& T_H\simeq \frac{1}{8\pi G_N m } \left(\, 1 +\frac{\alpha_q}{4m^2 G_N} e^{-4\mu m G_N}\,\right)\ .
\end{eqnarray}
The above results show only exponentially small deviations form the Schwarzschild black hole behavior. \\
In the opposite regime  $r_h < <1/2\mu$  ($m < <1/4G_N \mu$), there is no enough mass  to produce black hole and the source remains a point-particle.\\
In the intermediate case $r_h \sim 1/2\mu$ the horizon is dominated by quantum effects and becomes fuzzy.  An appropriate quantum description is needed.\\
One among various approaches is to mimic the  horizon fluctuations in terms of the equivalent oscillatory motion of
a  point-particle  in a suitably chosen potential \cite{Casadio:2013aua,Casadio:2015sda,Casadio:2016dzy}. 
In  recent  papers we proposed a  self-consistent way to obtain such a potential starting from the classical horizon equation itself\cite{Spallucci:2014kua,Spallucci:2016qrv,Spallucci:2021ljr}.  In the  present  model we follow the latter path and obtain the following relativistic wave equation
\begin{equation}
\frac{1}{r^2}\frac{d}{dr}\left[\,r^2 \frac{d\psi}{dr} \,\right]+\left[\,E^2-\frac{r^2}{4\, G_N^2}\left(\, 1 +\alpha_q\, G_N\,\frac{e^{-2\mu r}}{r^2 } \,\right)^2-\frac{l\,\left(\, l+1\,\right)}{r^2}\,\right]\psi\left(\, r\,\right)=0\ .
\label{weq}
\end{equation}
The wave function $\psi\left(\, r\,\right)$ represents the probability amplitude to find the horizon  at radial distance $r$ from the origin. $l$ is the angular momentum quantum number.  
At large distance $r> 1/2\mu$ the charge is completely screened and not accessible to an asymptotic observer. In this case one recovers the wave equation for a neutral Scharzschild black hole 

\begin{equation}
\frac{1}{r^2}\frac{d}{dr}\left[\,r^2 \frac{d\psi}{dr} \,\right]+\left[\,E^2-\frac{r^2}{4\, G_N^2}-\frac{l\,\left(\, l+1\,\right)}{r^2}\,\right]\psi\left(\, r\,\right)=0\ .
\label{weqSchw}
\end{equation}

Formula (\ref{weqSchw}) is nothing else but the equation for a relativistic quantum $3D$ harmonic oscillator with energy spectrum given by
\begin{eqnarray}
E^2_{n\,l} 
= \frac{1}{2G_N}\left[\, 4n+2l  + 3\,\right]\ ,\quad n=1\ ,2\ ,3\ ,\dots\ ,\quad l\le n-1\ .
\label{spectrum}
\end{eqnarray}

The presence of an electric charge affects  only the behaviour of the system below the screening length. In fact, for $r< 1/2\mu$ the exponential is close to one and the wave equation becomes

\begin{equation}
\frac{1}{r^2}\frac{d}{dr}\left[\,r^2 \frac{d\psi}{dr} \,\right]+\left[\,E^2- \frac{r^2}{4G_N^2}\left(\, 1 +\frac{\alpha_q G_N}{r^2}\,\right)^2-\frac{l\,\left(\, l+1\,\right)}{r^2}\,\right]\psi\left(\, r\,\right)=0
\label{weq3}
\end{equation}

In order to understand the physical meaning of the various terms in  (\ref{weq3})  it is useful to introduce the effective potential   $U_{eff}\left(\, r\,\right)$   as 

\begin{eqnarray}
&& \frac{1}{r^2}\frac{d}{dr}\left[\,r^2 \frac{d\psi}{dr} \,\right]+\left[\,E^2- \frac{\alpha_q}{2G_N}-U_{eff}\left(\, r\,\right)\,\right]\psi\left(\, r\,\right)=0\ ,\\
\label{weq4}
&&U_{eff}\left(\, r\,\right)\equiv \underbrace{\frac{r^2}{4G_N^2}}_{harmonic} + \underbrace{\frac{\alpha_q^2}{4r^2}}_{Coulomb}+\underbrace{\frac{l\left(\, l+1\,\right)}{r^2}}_{centrifugal}\ .
\label{effv}
\end{eqnarray}

The first term in (\ref{effv}) is the harmonic potential describing the Planck frequency oscillations of the quantum horizon;
the second term accounts for the charge \emph{ repulsive}  self-interaction; the third term is the well known centrifugal 
barrier.\\
The exact solution of the horizon wave-function reads:
\begin{equation}
 \psi\left(\,r\,\right)=N_n \left[\, \frac{r^2}{2G_N}\,\right]^s\, e^{-r^2/4G_N} L_n^{2s +1/2}\left(\, r^2/2G_N\,\right)\ ,
\end{equation}\vskip 10pt
where, $L_n^{2s +1/2}\left(\, \frac{r^2}{2G_N}\,\right)$  are generalized Laguerre polynomials; $N_n$ is the normalization coefficient, and the parameter $s\equiv \left[\, \sqrt{\alpha_q^2  +\left(\, 2l+1\,\right)^2 } -1\,\right]/2 $ 
As expected,  one finds a discrete energy spectrum for the quantum horizon given by
\begin{eqnarray}
E^2_{n\,l} 
= \frac{\alpha_q}{2G_N}+\frac{1}{2G_N}\left( 4n+2s  + 3+\alpha_q\right)\ ,
\label{spectrum}
\end{eqnarray}

and the ground state energy is:
\begin{equation}
E^2_{0\,0} =m^2_{Pl.}\left(2+\alpha_q\, +\sqrt{ \alpha_q^2  +1  }  \right)\ .
\end{equation}
We introduced the Planck mass as $m^2_{Pl.}\equiv 1/2G_N$. In the unit charge case $q=e$, $\alpha_e= e^2/4\pi \simeq 1/137$, $E^2_{0\,0}$ can be expanded  in powers of $\alpha_e$ and one finds 
\begin{equation}
E^2_{0\,0} =\left(3+\alpha_e \right)m^2_{Pl.}  +O\left(\, \alpha_e^2\,\right)\ .
\end{equation}

Let us also consider the spectrum of highly excited states  with $n>> 1\ , l >> 1$. In this limit $s \to l$ one has
\footnote{For a unit charge $\alpha_e$ is small but we keep it in the expansion in order  to appreciate first order correction with respect to  the neutral case.}
\begin{equation}
 G_N \,E^2_{n\,l}\simeq  2n+ l +\alpha_e\ . \label{largel}
\end{equation}

Equation (\ref{largel} ) allows  to express the angular momentum   $l$ as a "~function~"
of energy $E_n$. This behavior is characteristic of Regge trajectories used to describe strongly interacting resonances in
hadronic physics. In our case we find
\begin{eqnarray}
&& l \simeq \alpha^\prime \, E^2_{n\,l} + \beta_n\ ,\label{r1}\\
&& \alpha^\prime \equiv G_N=\frac{1}{2m^2_{Pl}}\ ,\label{r2}\\
&& \beta_n= -2n -\alpha_e\ .\label{r2}
\end{eqnarray}

Equation (\ref{r1}) describes a  \emph{family of trajectories}   $l=l\left(\,E^2\,\right) $ with  angular coefficient  $\alpha^\prime$ and intercept $\beta_n$. 
The importance of a relation like (\ref{r1}) is that naturally arises in  \emph{string theory}
where the Regge slope $\alpha^\prime$ is related to the \emph{string tension} $\rho$ through
the relation $\rho=1/2\pi \alpha^\prime $. 
Therefore, we arrive at the conclusion that higher excitations of the quantum horizon display a string-like nature.
The correct counting of the
black hole micro-states in agreement with the Area Law \cite{Susskind:1993ws} qualified String Theory as 
a legitimate candidate to account for quantum black holes. From this perspective highly excited strings shows 
thermodynamical properties analogous to those  of black holes
\cite{Horowitz:1996nw,Damour:1999aw,Veneziano:2004er}. 
In this paper we arrive the same conclusion following an inverse path: first we quantized the black hole horizon and then
discovered that highly excited states behave like strings.

\section{Discussion and conclusions}
\label{end}
In this paper we have reviewed the  double copy relationship connecting  gravitons and gauge fields. In its original version this correspondence was inspired by the structural similarity between corresponding perturbative scattering amplitudes in Yang-Mills theory and  General Relativity \cite{Bern:2019prr}. Once known in one case a simple "~recipe~" allows
to write the amplitudes in the other one without the need of lengthy calculations.\\
In the static limit one-boson exchange amplitudes gives the corresponding classical interaction potentials. In this way, one establishes a duality relation between Coulomb and Newton potentials. By implementing the unique properties of the Kerr-Schild coordinate system one reaches the conclusion that a similar duality relation connects the Coulomb field and the Scharzschild metric despite the absence of any electric charge in the latter. It follows that it is not immediate to accommodate  into this framework  charged solutions of the Einstein equations such as the Reissner-Nordstr\"om one and a non-perturbative approach is necessary.\\ 
The fundamental feature of perturbative  double copy is to express the tensor structure of a vector gauge potential and the graviton in terms of the same null vector $l_\mu$.  The only distinction between the two interactions is encoded into
different scalar potentials. In summary:\\
\begin{center}
Perturbartive  double copy 
\begin{eqnarray}
&& A_\mu = \phi_{gauge}\, l_\mu\ ,\label{p1}\\
&&\kappa h_{\mu\nu}= -2 \phi_{grav}\, l_\mu l_\nu\ . \label{p2}
\end{eqnarray}
\end{center}
\begin{center}
 General Relativity 
\begin{equation}
A_\mu = \phi_{gauge}\, l_\mu\ ,\label{gr1}
\end{equation}
 \begin{equation}
 \kappa h_{\mu\nu}= -2 \phi_{grav}\, l_\mu l_\nu + 8\pi G_N\phi^2_{gauge}\, l_\mu \, l_\nu\ . \label{gr2}
\end{equation}
 \end{center}
 Notice that $\phi_{grav}$ is the same both in (\ref{p2}) and (\ref{gr2}). However, in the perturbative approach it is obtained through a substitution rule
 between coupling constants while in General Relativity it represents the solution of the homogeneous Poisson equation.
 The second term in (\ref{gr2}) is the contribution from the energy density stored in the electric field which is missing in (\ref{p2}). It enters equation (\ref{tre}) through the energy-momentum tensor. \\
 Let us remark that the form (\ref{peq})  of $T^0{}_0$ is a \emph{sufficient condition} to
 obtain a NDC-type solution for the gravitational field.
 We checked that (\ref{gr2}) holds in Einstein-Maxwell theory even if the gauge symmetry is spontaneously broken.
 In this case the electric field contribution is damped due to the presence of the photon mass in the Higgs true vacuum. 
 We  considered the resulting screened Reissner-Nordstr\"om black hole both in the "~large~" and "~small~" limits.
  When the horizon radius and the massive photon Compton wavelength become comparable a quantum description is
  needed.   We determined the energy
 spectrum for this type of object and found interesting analogy with the behaviour of highly excited strings.

\end{document}